\documentclass{PoS}
\usepackage{units}

\title{Livetime and sensitivity of the ARIANNA Hexagonal Radio Array}

\ShortTitle{Livetime and sensitivity of the ARIANNA Hexagonal Radio Array}

\author{\speaker{Anna Nelles} and Chris Persichilli for the ARIANNA Collaboration\\
        Department of Physics and Astronomy, University of California Irvine\\
        Irvine, CA 92697-4575, USA\\
        E-mail: \email{anelles@uci.edu}}

\abstract{The ARIANNA collaboration completed the installation of the hexagonal radio array (HRA) in December 2014, serving as a pilot program for a planned high energy neutrino telescope located about 110 km south of McMurdo Station on the Ross Ice Shelf near the coast of Antarctica.  The goal of ARIANNA is to measure both diffuse and point fluxes of astrophysical neutrinos at energies in excess of  $\unit[10^{16}]{eV}$. Upgraded hardware has been installed during the 2014 deployment season and stations show a livetime of better than 90\% between commissioning and austral sunset. Though designed to observe radio pulses from neutrino interactions originating within the ice below each detector,  one station was modified to study  the low-frequency environment and signals from above. We provide evidence that the HRA observed both continuous emission from the Galaxy and a transient solar burst.  Preliminary work on modeling the (weak) Galactic signal confirm the absolute sensitivity of the HRA detector system.}

\FullConference{The 34th International Cosmic Ray Conference,\\
		30 July- 6 August, 2015\\
		The Hague, The Netherlands}

\begin{document}

\section{Introduction}

\begin{figure}
\includegraphics[width=.49\textwidth]{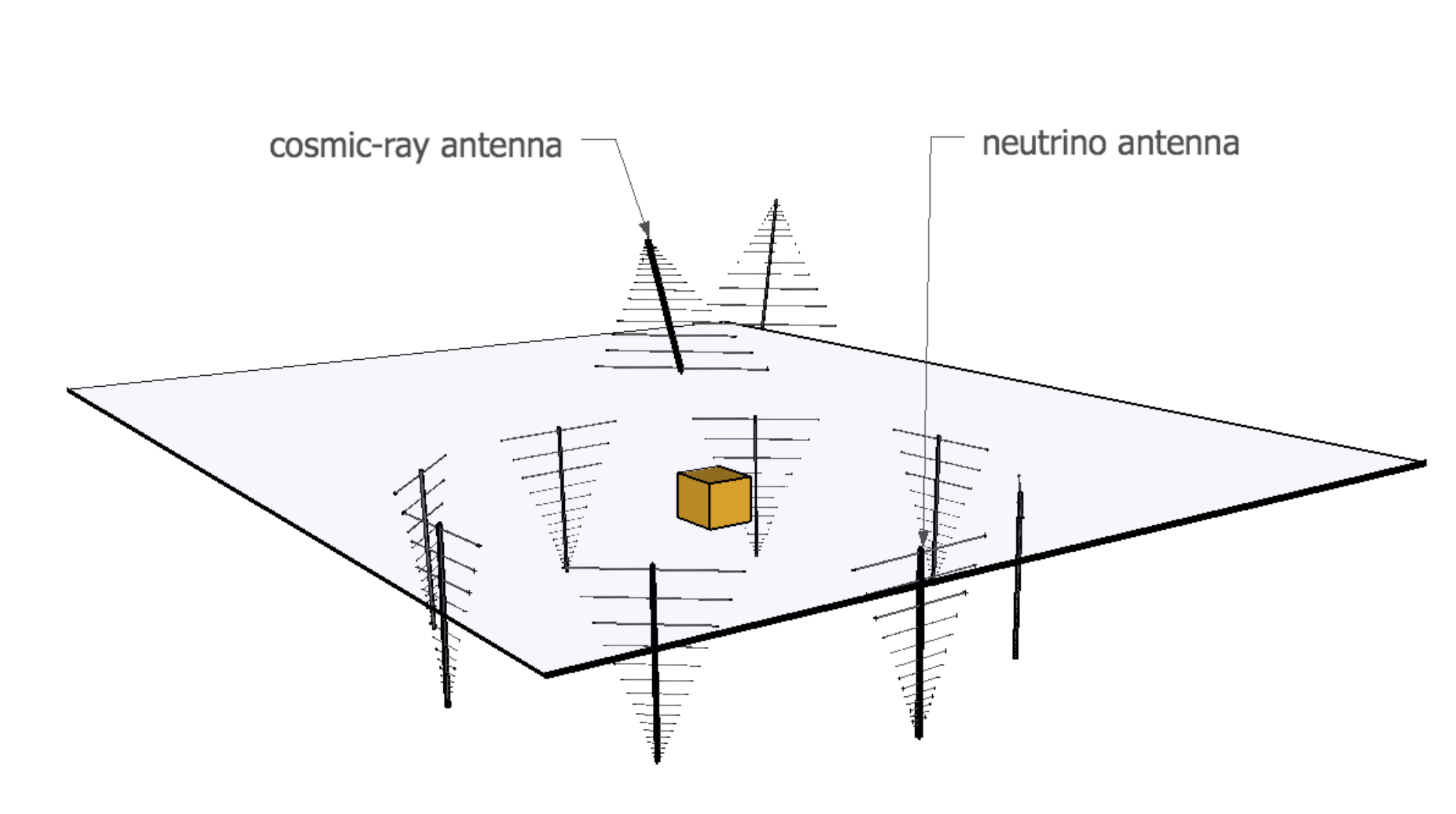}
\includegraphics[width=.49\textwidth]{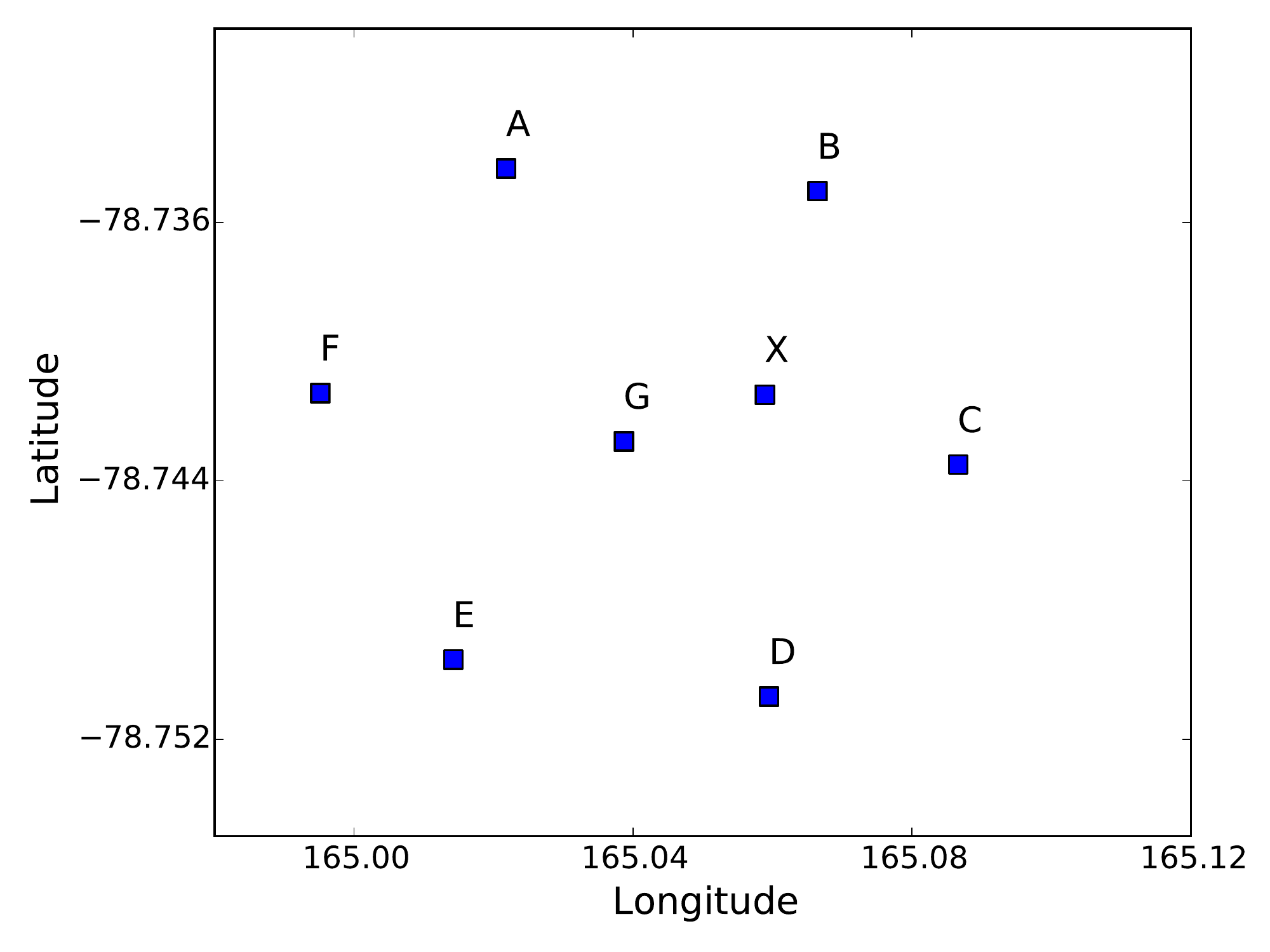}
\caption{Left: Schematic of straw man design for an ARIANNA station, with 8 downward facing LPDA receivers and two upward tilted LPDA receivers.  Most HRA stations consist of only 4 downward facing and no upward facing antennas, except station X, which has two LPDAs tilted upward from horizontal by 30 degrees, with one nose pointing north and the other south. Right: Geographic coordinates (in Latitude and East Longitude) of the stations that comprise the HRA detector. }
\label{layout}
\end{figure}

The ARIANNA experiment is designed to detect diffuse, unresolved, and point fluxes of neutrino emission in the energy interval between $\unit[10^{8}-10^{10}]{GeV}$ \cite{2015ARIANNALimitsPaper}, and work has started to investigate its capabilities down to $\unit[10^7]{GeV}$. The primary objective of ARIANNA is to measure neutrinos associated with the flux of high energy protons and their interaction with the cosmic microwave background \cite{Greisen1966,Zatsepin1966,Stecker1973}.  The ARIANNA collaboration plans to construct a surface array of independent, autonomous stations encompassing more than $\unit[1000]{km^2}$ on the Ross Ice Shelf in Antarctica.  The hexagonal radio array (HRA) serves as a pilot program for ARIANNA \cite{2015ARIANNATechnologyPaper}. Each HRA station consists of four downward facing log-periodic dipole antennas (LPDA), though the baseline design for the next stage of development consists of 6-8 downward facing LPDAs (Figure \ref{layout}). These antennas will be sensitive to the radio emission that is created by the interaction of a neutrino in the ice. The ice to water interface at the bottom of the ice acts as a mirror that reflects the radio signals to the antennas at the surface \cite{Glaciology}. This paper describes the completed HRA, and its operation and  performance between commissioning in mid-December 2014 and Austral sunset in mid-April 2015.  A report on the updated search for neutrinos can be found in \cite{ReedICRC2015}.

\section{Array layout and station hardware}

The right panel of Figure \ref{layout} shows the geographic location (in latitude and east longitude) of each detector station of the HRA.  Each site is a fully autonomous detector, instrumented with four LPDAs buried just below snow level. All necessary components of the detector station, including renewable power, data acquisition, and communications, require about 4 hours to install. 
The aluminium LPDAs have a boom length of $\unit[1.4]{m}$ and are designed for frequencies (in air) of $\unit[105-1300]{MHz}$. Each station includes signal amplification spanning the $\unit[0.1-1]{GHz}$ bandwidth, 2 G-sample/s data acquisition and digitization, a multi-level trigger, computing hardware including 32 GB of solid-state data storage, system control and monitoring, a solar power tower and battery backup. Two communication modalities have been incorporated: The stations can receive commands and transmit data by long-distance wireless via a repeater, or they can do so by Iridium short-burst data messaging. For the latter, the bandwidth is limited to about 300 bytes per message, but on-site tests of low trigger-rate configurations have shown that this can be sufficient for real-time data transmission. The Iridium communication consumes less power and does not rely on a relay-tower.
Two generations of ARIANNA station electronics have been deployed in two stages. Three HRA sites, plus site X, contain the first \emph{ATWD}  generation as discussed in  \cite{2015ARIANNATechnologyPaper}. Four newer stations are based on the  \emph{SST} data acquisition chip and consist of a single data acquisition board, as opposed to a motherboard plus daughter-card arrangement. They are considerably less expensive, use one third the power, offer deeper analog waveform capture (4 channels of 256 samples at 2 G-samples/s per chip) and include a simplified yet high-performance trigger system \cite{Kleinfelder2015}. The updated stations also include improved amplifiers with flatter frequency response, greater stability and the integration of all band-pass filtering.

\section{Array performance}
The current array has been operational between December 2014 and April 2015, limited by winter's permanent night on the Ross Ice Shelf.

\subsection{Rates and Livetimes}

\begin{figure}
\includegraphics[width=1.\textwidth]{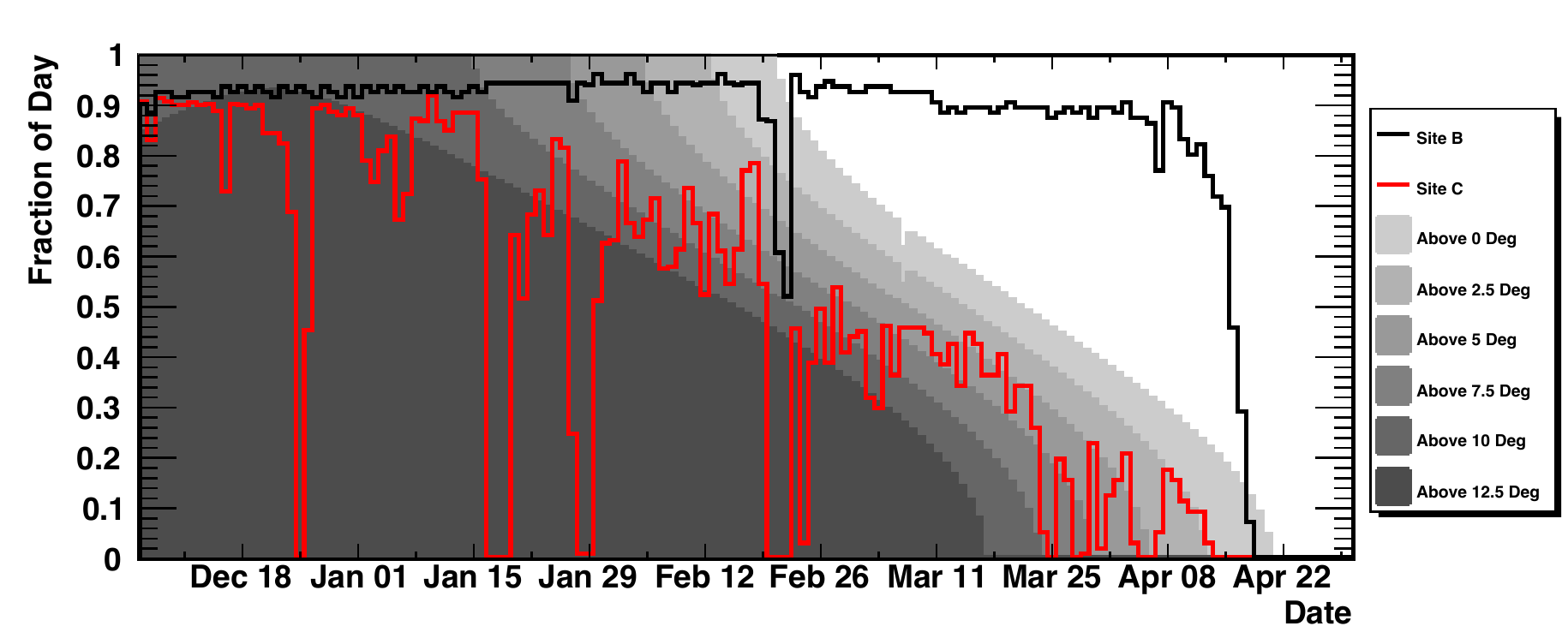}
\caption{Fraction of day in which stations are operational as a function of time. Shown are examples of a station with functional battery (black line) and without (red line). The shaded regions in the background indicate the fraction of a day in which the Sun was higher than a certain angle above the horizon. The Sun sets for Antarctic winter on April 20th.}
\label{livetime}
\end{figure}

Figure \ref{livetime} shows the fraction of a day in which a station was operational. Two stations are shown, one with a battery and one in which the battery failed. The station with battery consistently shows an uptime of more than 90\%. This station was up and taking data for a total of 121 days during this season. The deadtime is typically less than 6\%, which is primarily determined by the length of communications during which the data-taking is paused (about one minute every half hour). The collection of calibration data (typically ten minutes every twelve hours) is also included as livetime, though it is currently not used in the data analysis. The only significant dip visible in Figure \ref{livetime} is a brief period in February, in which a heavy storm was recorded at the ARIANNA site. This caused repeated failures of long-distance wireless communication, and led to increased deadtime on many stations. During this period, some data was successfully transmitted from Site B in real-time using the Iridium network. In general, the station remains at high efficiency until the Sun sets for the Antarctic winter. The station without functional battery shows the feature of decreasing livetime as a function of daylight, as well as a greater number of inefficient periods. This particular station suffered from a software failure that lead to occasional multi-day outages. This issue was investigated, and the patch remotely distributed to the ARIANNA stations, resolving the issue in February 2015. It is interesting to note that the power delivered by the Sun is already sufficient once it is more than $2^{\circ}$ above the horizon. However, a station without battery is more sensitive to transient weather conditions, such as heavy storms like the one discussed above. These results clearly illustrate the positive effect of a battery on livetime. Testing of a new generation of batteries that are better able to withstand the cold temperatures is currently underway, and yielding promising results. All stations will be equipped with batteries in the next season. Our measured consumption of 6 Watts or less per station means that a relatively small battery pack can dramatically improve station livetime. 
During the operational season the stations triggered at rates of less than 0.1 Hz at thresholds of $4\sigma$, which is a clear indication for the suitability of the location of ARIANNA and the absence of human-made radio-interference. Exceptions to this low trigger rate can be found during heavy storms and due to astronomical influences (see below). 

\subsection{Sensitivity to background of Galactic and solar radio emission}
\begin{figure}
\includegraphics[width=0.49\textwidth]{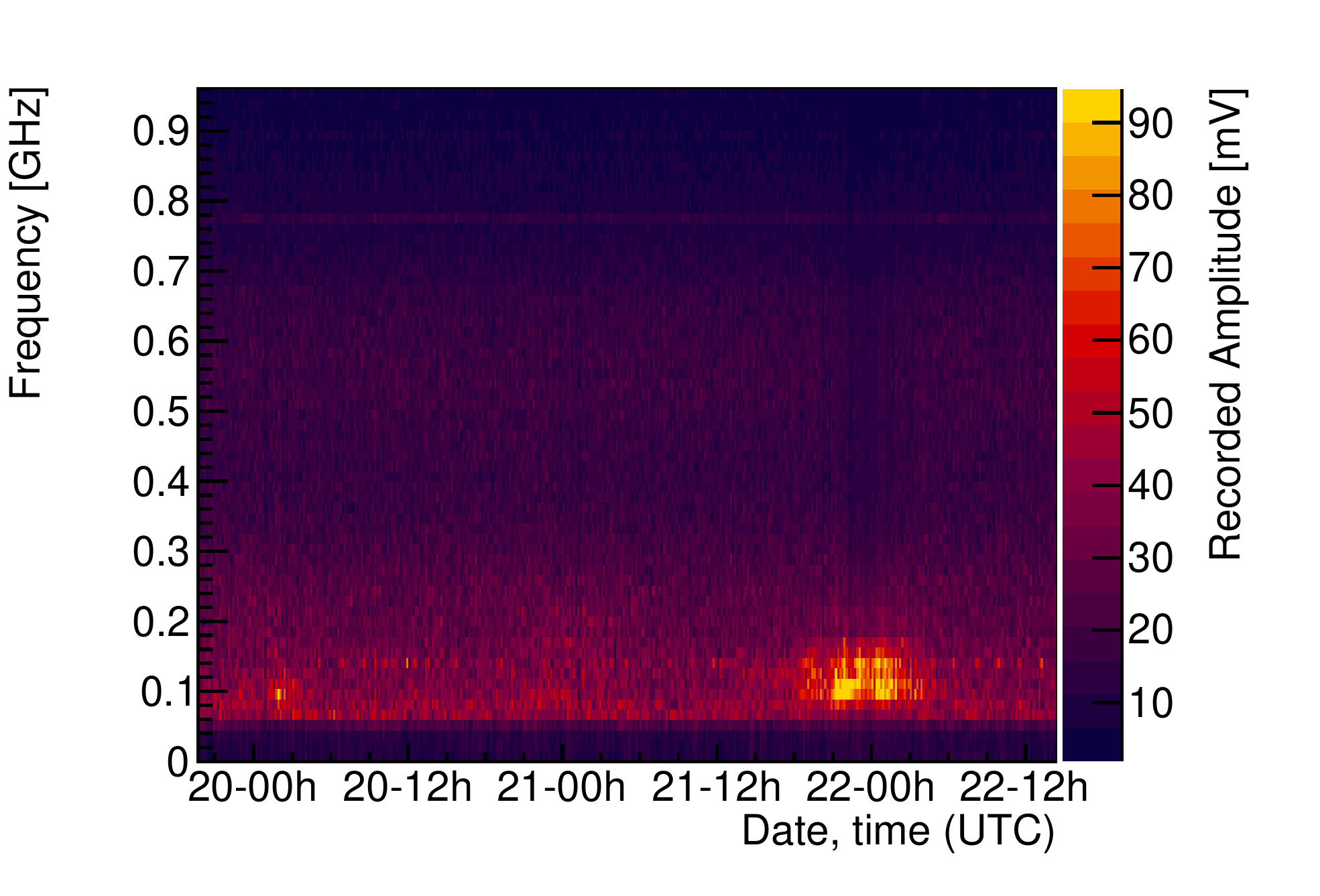}
\includegraphics[width=0.49\textwidth]{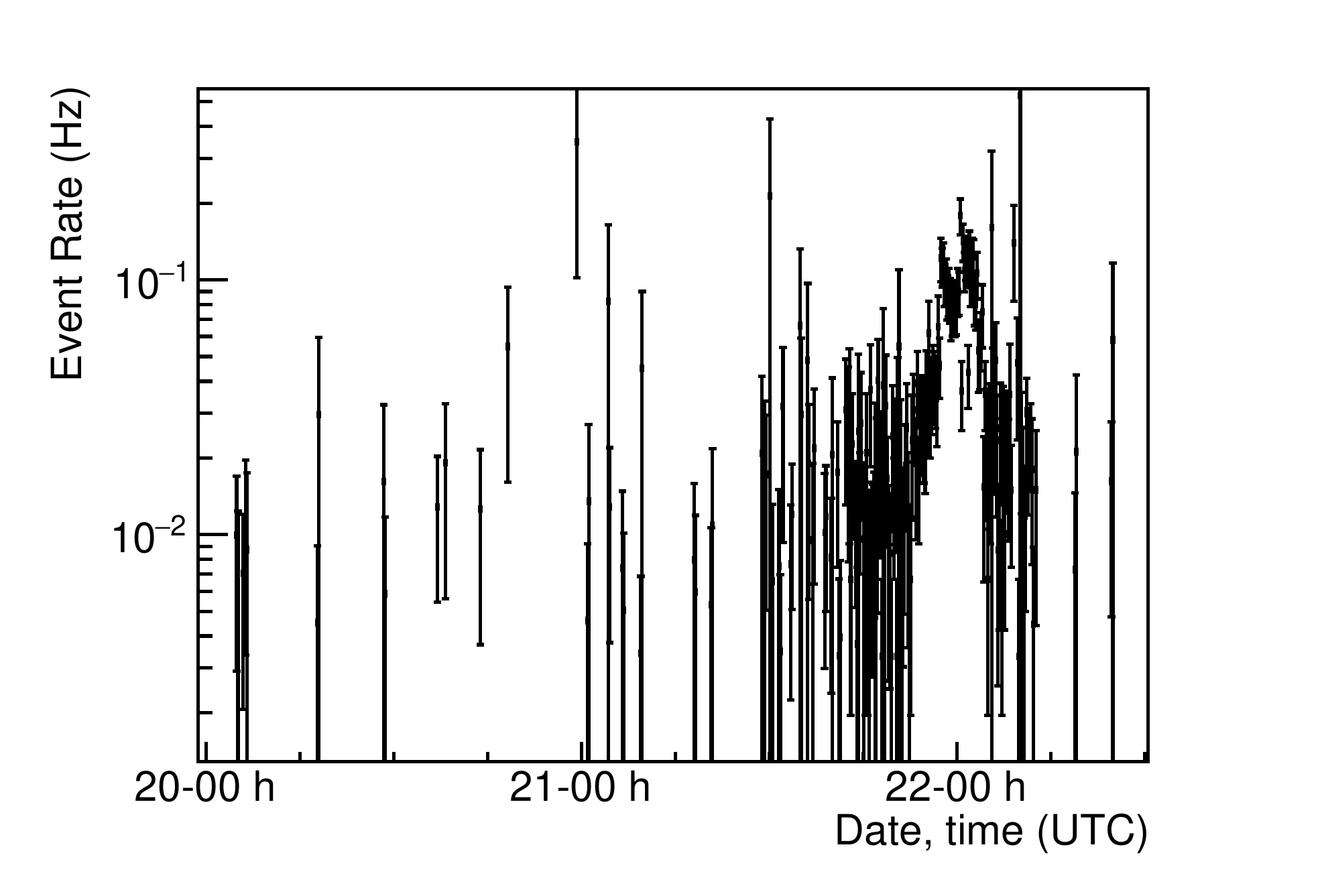}
\caption{Left: Measured amplitude spectrum as function of time  in December 2014 in one upward pointing antenna. An increase is visible at 1:00 AM on December 20th, which coincides with a coronal mass ejection, observed in X-ray and radio data.  The Sun stays radio-loud after this burst and its transient can be observed, especially obvious on the 22nd, which coincides with increased auroral activity. The solar modulation is not visible in earlier data. Right: Trigger rates as function of time in the downward pointing antenna. The solar-loud phase of the Sun, results in increased trigger rates, which indicates that the Sun has also been detected in the downward facing antennas. }
\label{solar}
\end{figure}

At frequencies below  $\unit[\sim150]{MHz}$ the radio emission of the Galaxy contributes a significant fraction to the noise measured with well-designed antenna systems. The radio emission from the Sun will only be visible with a single antenna in periods where the Sun is \emph{radio-loud}. This refers to \emph{solar bursts} of various types, linked to the ejection of material from the surface of the Sun. 
A very strong solar burst occurred in the night of December 20th, 2014 \cite{SpaceWeather}, causing very strong X-ray flares, coronal mass-ejection, and strong auroras on Earth during the days following the burst. Signals coinciding with this burst and the auroras were found with high significance in the upward pointing antennas of the station at position X (Figure \ref{solar}). In addition, a significant increase in the trigger rates of the downward pointing antennas was found. This is the only significant increase that has been found during good weather conditions in the 2014/15 season.

\begin{figure}
\centering
\includegraphics[width=0.6\textwidth]{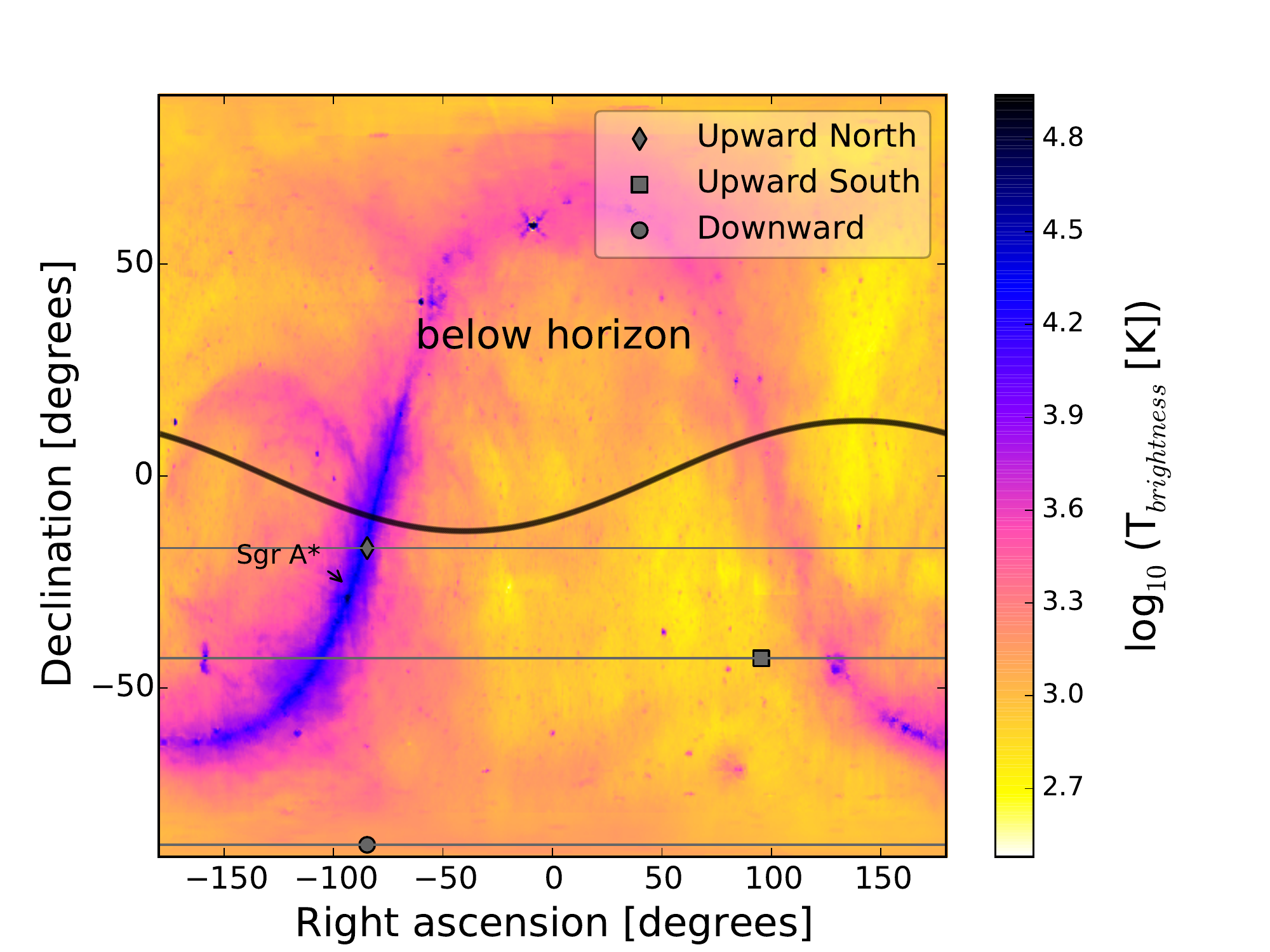}
\caption{Model of the radio emission (brightness temperature) at 85 MHz as derived from LFMap \cite{LFmap} as function of right ascension and declination on the sky. Also indicated is the horizon (thick black line) at a given time and the trajectories of the three different antenna pointings on the sky as markers with lines. These models are generally believed to have a scale uncertainty of 15\%.}
\label{galaxy}
\end{figure}

If the Sun is is radio-quiet, the diffuse radio emission of the Galaxy becomes the most prominent background signal below $\unit[\sim150]{MHz}$. As the radio emission of the Galaxy falls off exponentially with frequency, the largest contributions are expected at the lowest detectable frequencies. As the contribution of the emission to the thermal noise of the system is estimated to be small, a stacking needs to be performed to identify the signature. As the signal is periodic in local sidereal time (LST), in which one LST day is about 23 hours and 56 minutes, the amplitude measured in the lowest relevant frequency bins is stacked in LST bins. This amplifies signals periodic in LST and smears out any signals periodic in local time. Any significant variation periodic in LST is therefore evidence for the sensitivity of the system to the Galactic emission. 
Models of the radio emission, such as LFMap \cite{LFmap} (see Figure \ref{galaxy}), can be used in combination with a model of the antenna sensitivity \cite{2015ARIANNATimeResponsePaper} and the amplifier to predict the amount of Galactic emission that is measured. The ARIANNA LPDAs are less sensitive below 100 MHz due to their dimensions. However, it is likely that the antenna behavior in the top layer of the ice firn changes such that it will be more sensitive to lower frequencies, since the speed of light in the medium is decreased \cite{ARIANNAPrototype}. The maximum power is indeed observed around $\unit[85]{MHz}$, which confirms that the antenna response is affected by the medium. In addition, Fresnel bending was taken into account when computing the arrival direction of the signal, but this did not affect the amplitude or time dependence appreciably. 

\begin{figure}
\centering
\includegraphics[width=0.6\textwidth]{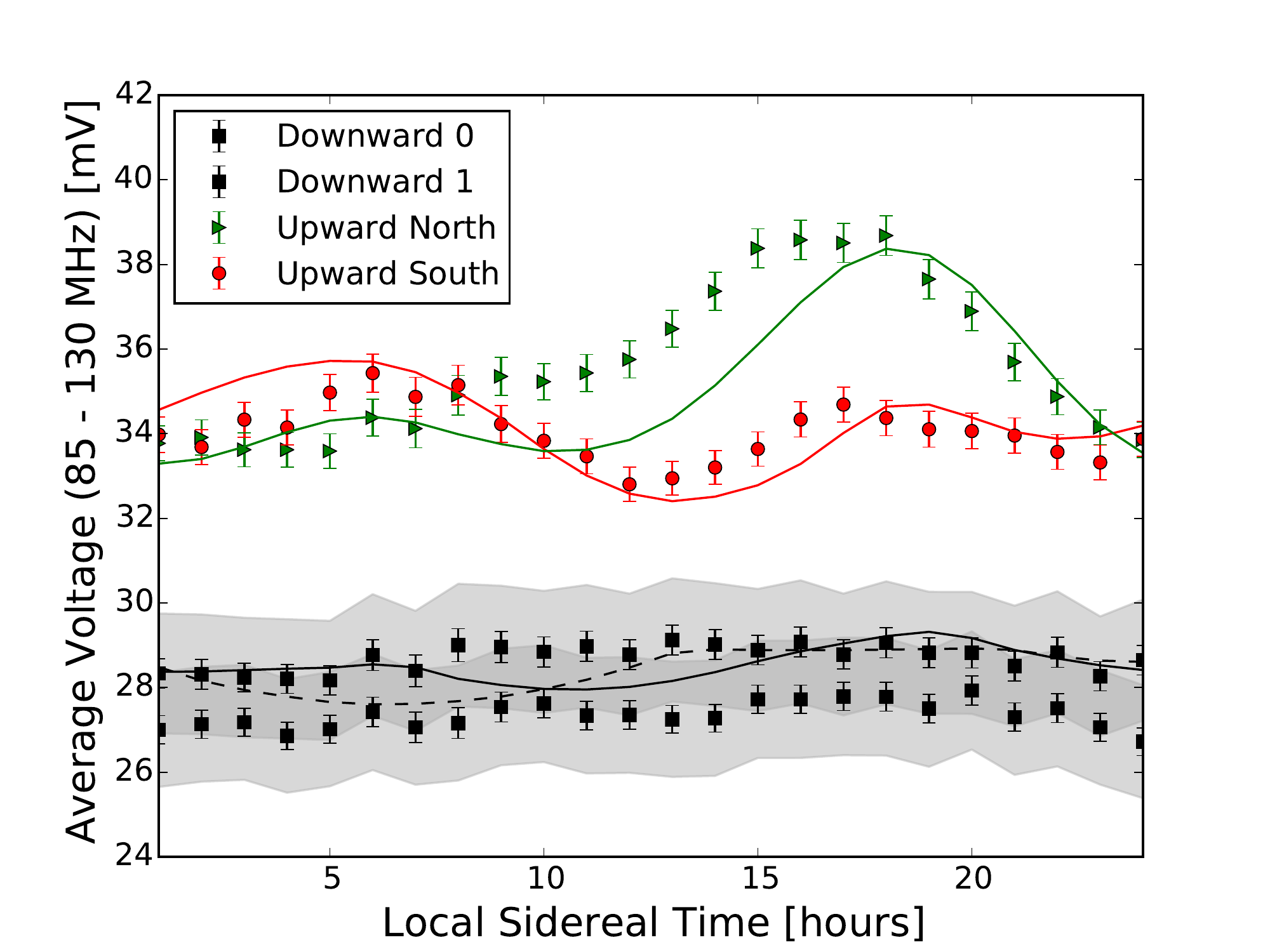}
\caption{Measured and modelled amplitude in the Fourier Transform in the range of 85-130 MHz as function of local sidereal time. Shown are the measurements (points) and the predicted amplitudes (lines) for four types of antennas. Due to a missing calibration on the downward facing antennas of station X, an additional uncertainty is plotted.}
\label{galaxy_data}
\end{figure}

A measurement of this Galactic variation is shown in Figure \ref{galaxy_data}, using data from the station at location X, which is equipped with upward tilted antennas. The Figure shows the average voltage amplitude measured in forced triggers in the bins of the Fourier transform between 85 and 130 MHz as a function of LST. Overlaid is a model of the Galactic emission. The power from the Galactic emission has been summed with the power from the intrinsic system noise as measured, to obtain the complete prediction as shown in Figure \ref{galaxy_data}.
It is shown in the figure that the downward pointing antennas should detect very little variation in their back-lobe, as already indicated in Figure \ref{galaxy}. Also, almost no difference is expected between differently oriented channels. This is confirmed in data. The upward pointing antennas are very sensitive to the Galactic emission, as the front-lobe of the LPDA has the largest sensitivity and filters below 100 MHz have been removed on the upward facing channels. A significant variation is detected. The shape of the prediction does not fully match the measurements. It should be noted that the predictions are very sensitive to the antenna model used  and an in-depth study of the effects of the ice on the antenna behavior is currently underway. However, these first predictions confirm the general antenna behaviour, as well as the measured amplitudes, which is an essential step in the understanding of the hardware.

\section{Background of cosmic rays}
In the radio-quiet environment at the ARIANNA site, the largest background, if any,  is expected to be caused by air showers induced by cosmic rays. Air showers show the strongest signals at lower MHz frequencies. On a ring at the Cherenkov-angle higher frequencies are also observed. However, the air shower radio footprint is largest at frequencies below 100 MHz, which is at the edge of the sensitivity of the ARIANNA band. The threshold energy for cosmic ray detection is therefore determined in a region where small uncertainties in the modeling of the antenna have a large impact. In addition, air showers arrive in the less sensitive and less well-defined back-lobe of the antenna. 
To study the threshold energy and the expected rates in further detail, a library of 600 CoREAS air shower simulations \cite{Coreas} (CORSIKA 7.4005, QGSJET-II-04, FLUKA) has been generated. The simulated showers cover the energy range of $\unit[10^{16}-10^{20}]{eV}$. Using proton primaries covers the full range of $X_{\mathrm{max}}$ more easily. Since the size of the radio footprint scales almost exclusively with the distance to the shower maximum, this will allow us to reweigh showers according to different composition assumptions later in the analysis. 
Previous studies show that the trigger rates depend sensitively on the accuracy of the LPDA response to low frequencies in the backward lobe direction.  We plan to combine the low frequency response obtained in the previous section with improved modeling of the backward lobe in mixed media to obtain an improved estimate of the cosmic ray trigger rates.  Simultaneously, we have started to develop new analysis procedures to specifically search for cosmic ray events in HRA data.   

\begin{figure}
\centering
\includegraphics[width=0.6\textwidth]{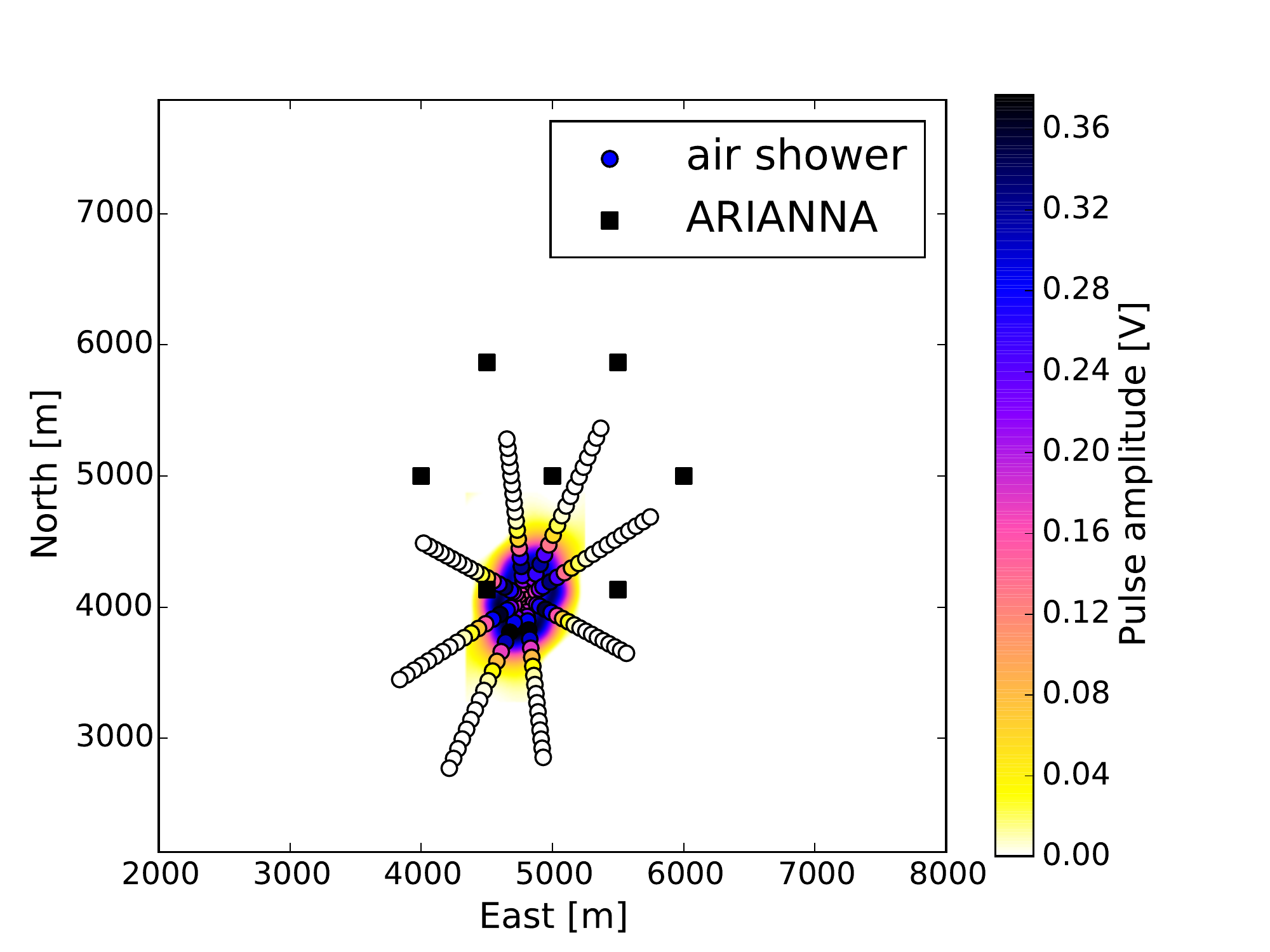}
\caption{Example of an air shower simulation as performed for ARIANNA. The radio signals are simulated on a star-shaped pattern \cite{Buitink2014}. This allows for the interpolation of the signals, which in turn allows us to choose different array layouts and shower core positions. The predicted signals are a combination of CoREAS simulations and the models of the ARIANNA hardware. }
\label{mc_CR}
\end{figure}

\section{Conclusions and Outlook}
We have presented the performance of the HRA, the pilot program for the ARIANNA high energy neutrino detector. During austral summer, the stations show a livetime of more than 90\% when equipped with a battery to supply power during March and April when the Sun periodically dips below the horizon. This is compatible with previous livetime measurements from the HRA-3. The upgraded data-acquisition system has reduced the power consumption to less than 6W,  and doubled the waveform record to 256 samples. Most data was transferred in realtime by high-speed wireless links to McMurdo Station, and the relatively small fraction of remaining data will be collected during the next Antarctic campaign.  In addition, the trigger rates were temporarily reduced to $\unit[\sim]{mHz}$ to test complete data transfer by SBD Iridium satellite modem with encouraging results. The observation of continuous Galactic emission was used to demonstrate that the detector sensitivity is understood in the frequency range below $\unit[150]{MHz}$, which is an important frequency band for both neutrino and cosmic ray flux estimates. It is expected that HRA will resume routine operation in late September, 2015 when sunlight is again sufficient to power the stations.           
 
 \section{Acknowledgements}
We thank the staff of Antarctic Support Contractors, Lockheed, and the entire crew at McMurdo Station for excellent logistical support. This work was supported by funding from the Office of Polar Programs and Physics Division of the US National Science Foundations, grant awards ANT-08339133, NSF-0970175, and NSF-1126672.  A. Nelles is supported by a research fellowship of the German Research Foundation (DFG), grant NE 2031/1-1.

\bibliographystyle{JHEP}
\bibliography{BIB}

\providecommand{\href}[2]{#2}\begingroup\raggedright\begin{thebibliography}{10}

\bibitem{2015ARIANNALimitsPaper}
{\bf ARIANNA} Collaboration, S.~W. {Barwick} et~al., {\it {A first search for
  cosmogenic neutrinos with the ARIANNA Hexagonal Radio Array}},  {\em
  Astroparticle Physics} {\bf 70} (Oct., 2015) 12--26.

\bibitem{Greisen1966}
K.~Greisen {\em Physical Review Letters} {\bf 16} (1966) 748.

\bibitem{Zatsepin1966}
G.~Zatsepin and V.~Kuz'min {\em JETP Letters} {\bf 4} (1966) 78.

\bibitem{Stecker1973}
F.~W. {Stecker}, {\it {Ultrahigh Energy Photons, Electrons, and Neutrinos, the
  Microwave Background, and the Universal Cosmic-Ray Hypothesis}},  {\em
  Astrophysics and Space Science} {\bf 20} (Jan., 1973) 47--57.

\bibitem{2015ARIANNATechnologyPaper}
{\bf ARIANNA} Collaboration, S.~W. {Barwick} et~al., {\it {Design and
  Performance of the ARIANNA HRA-3 Neutrino Detector Systems}},  {\em IEEE
  Transactions on Nuclear Science (in press)} (2015).

\bibitem{Glaciology}
{\bf ARIANNA} Collaboration, J.~C. {Hanson} et~al., {\it Radar absorption,
  basal reflection, thickness and polarization measurements from the ross ice
  shelf, antarctica},  {\em Journal of Glaciology} {\bf 61} (2015).

\bibitem{ReedICRC2015}
{C. Reed for the ARIANNA Collaboration}, {\it {Performance of the Completed
  ARIANNA Hexagonal Radio Array}},  {\em These Proceedings} (2015).

\bibitem{Kleinfelder2015}
S.~A. Kleinfelder, E.~Chiem, and T.~Prakash, {\it {The SST Fully-Synchronous
  Multi-GHz Analog Waveform Recorder with Nyquist-Rate Bandwidth and Flexible
  Trigger Capabilities}},  {\em Proc. IEEE Nuclear Science Symposium, Seattle,
  WA} (2014).

\bibitem{SpaceWeather}
{NOAA, Space Weather Prediction Center}, 2014.

\bibitem{LFmap}
E.~Polisensky, {\it {LFmap: A Low Frequency Sky Map Generating Program.}},
  {\em Long Wavelength Array (LWA) Memo Series} {\bf 111} (2007).

\bibitem{2015ARIANNATimeResponsePaper}
{\bf ARIANNA} Collaboration, S.~W. {Barwick} et~al., {\it {Time-domain response
  of the ARIANNA detector}},  {\em Astroparticle Physics} {\bf 62} (Mar., 2015)
  139--151.

\bibitem{ARIANNAPrototype}
L.~Gerhardt, S.~Klein, T.~Stezelberger, S.~Barwick, K.~Dookayka, et~al., {\it
  {A prototype station for ARIANNA: a detector for cosmic neutrinos}},  {\em
  Nucl.Instrum.Meth.} {\bf A624} (2010) 85--91.

\bibitem{Coreas}
T.~{Huege}, M.~{Ludwig}, and C.~W. {James}, {\it {Simulating radio emission
  from air showers with CoREAS}},  in {\em American Institute of Physics
  Conference Series}, vol.~1535 of {\em American Institute of Physics
  Conference Series}, pp.~128--132, May, 2013.

\bibitem{Buitink2014}
S.~{Buitink} et~al., {\it {Method for high precision reconstruction of air
  shower X$_{max}$ using two-dimensional radio intensity profiles}},  {\em
  Physical Review D} {\bf 90} (Oct., 2014) 082003.

\end{thebibliography}\endgroup

\end{document}